      [1999/12/01 v1.4c Il Nuovo Cimento]
\documentclass[varenna]{cimento}
\usepackage{amsmath}
\usepackage{amssymb}
\usepackage{citesort}
\usepackage{graphicx}
\def\ep{\varepsilon_p}
\title{Spectral signatures of Holstein polarons}
\author{Holger Fehske \atque Andreas Alvermann}
\institute{Institut f\"ur Physik, 
Ernst-Moritz-Arndt-Universit\"at Greifswald,
17487 Greifswald, Germany}
\author{Martin Hohenadler}
\institute{Institut f\"ur Theoretische Physik -- Computational Physics, 
TU Graz, 8010 Graz, Austria}
\author{Gerhard Wellein}
\institute{Regionales Rechenzentrum Erlangen, Universit\"at Erlangen, 91058
  Erlangen, Germany}
\shortauthor{H. Fehske, A. Alvermann,  M. Hohenadler \atque G. Wellein}
\begin{document}
\maketitle
\section{Fundamentals}
\subsection{Self-trapping phenomenon}
\begin{figure}[t]
\begin{minipage}{0.6\linewidth}
\includegraphics[width=.9\linewidth,clip]
{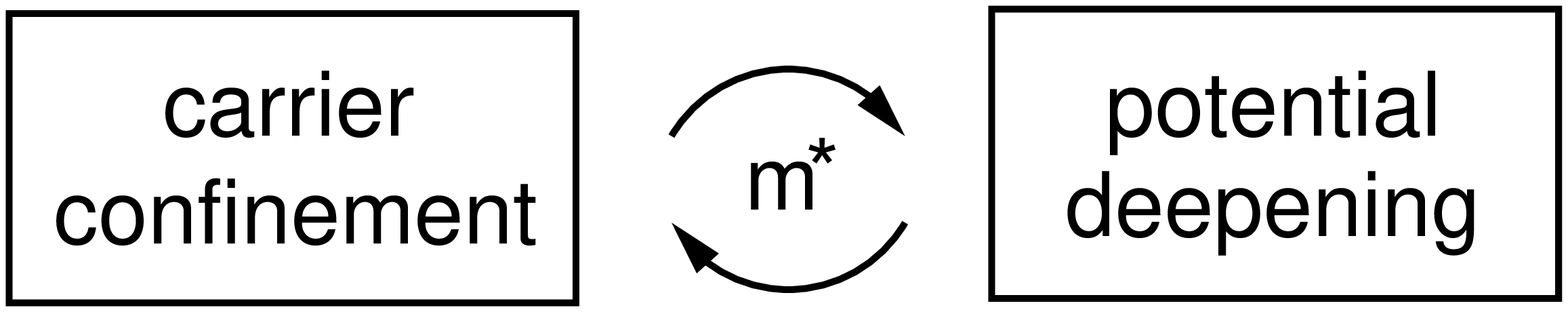}
\end{minipage}\hspace*{0.5cm}
\begin{minipage}{0.35\linewidth}
\caption{Polaron formation: The standard phase transition concept
fails to describe this highly non-linear feedback phenomenon.\vspace*{0.2cm}
\label{st}}
\end{minipage}
\end{figure}

Electrons or holes delocalised in a perfect rigid lattice can be ``trapped''
in a potential well produced by displacements of atoms if the
particle-lattice interaction is sufficiently strong~\cite{Fi75}.  Such
trapping is energetically favoured over wide-band Bloch states if the
carrier's binding energy exceeds the strain energy required to produce the
trap. Since the potential itself depends on the carrier state, this highly
non-linear process is called ``self-trapping'' (Fig.~\ref{st}).  A
self-trapped state is referred to as ``large'' if it extends over multiple
lattice sites. Alternatively, for a quasi-particle (QP) with
extremely large effective mass $m^*$ -- practically confined to a single
site -- the state is designated as ``small''.  Self-trapping does not imply a
breaking of translational invariance, i.e., in a crystal these eigenstates
are still itinerant allowing, in principle, for coherent transport with an
extremely small bandwidth.
 
Introducing the concept of polarons into physics, the possibility of electron
self-trapping was pointed out by Landau as early as 1933~\cite{La33}.
Self-trapped polarons consisting of electrons accompanied by phonon clouds
can be found, e.g., in alkali(ne) metal (earth) halides, II-IV- and group-IV
semiconductors, and organic molecular crystals~\cite{SS93}.  With the
observation of polaronic effects in high-$T_c$ cuprates~\cite{AM94} and
colossal magneto-resistance manganites~\cite{JHSRDE97}, research on polarons
has attracted renewed attention (see also~\cite{Ra_Varenna}).

\subsection{Holstein model}
Depending on the relative importance of long- and short-range
electron-lattice interaction, simplified models of the Fr\"ohlich~\cite{Fr54}
or Holstein~\cite{Ho59a} type have been widely used to analyse polaronic
effects in solids with displaceable atoms. The spinless Holstein 
Hamiltonian reads
\begin{equation}
H=-t\sum_{\langle i,j\rangle} c_i^{\dagger} c_j^{}  -\sqrt{\ep \omega_0}\sum_i
(b_i^{\dagger}  + b_i^{})  n_i^{}
+\omega_0 \sum_i  b_i^{\dagger} b_i^{}\,.
\label{hm}
\end{equation}
Here $ c_i^{[\dagger]}$ and $ b_i^{[\dagger]}$ are the annihilation
[creation] operators of a (spinless) fermion and a phonon at Wannier site
$i$, respectively, and $n_i^{}=c_i^{\dagger}c_i^{}$. 
In~(\ref{hm}), the following idealisations of real
electron-phonon (EP) systems are made: (i) the electron transfer $t$ is
restricted to {\it nearest-neighbour} pairs $\langle ij\rangle$; (ii) the
charge carriers are {\it locally} coupled to a {\it dispersionless} optical
phonon mode [$\ep$ measures the polaron binding energy and $\omega_0$ is the
bare phonon frequency ($\hbar=1$)]; (iii) the phonons are treated within the
{\it harmonic} approximation.

Nevertheless, as yet, none of the various analytical treatments, based on
weak- and strong-coupling adiabatic and anti-adiabatic perturbation
expansions~\cite{Mi58}, are suited to investigate the physically
most interesting polaron transition region.  In the latter, the
characteristic electronic and phononic energy scales are not well separated
and non-adiabatic effects become increasingly important, implying a breakdown
of the standard Migdal approximation.  Quasi-approximation-free numerical
methods like quantum Monte Carlo (QMC)~\cite{RL83,HNLWLF05,Mi_Varenna}, exact
diagonalisation (ED)~\cite{RT92}, or the density matrix renormalisation group
(DMRG)~\cite{JW98a} can, in principle, close the gap between the weak- and
strong-EP-coupling limits, therefore representing the currently most reliable
tools to study polarons close to the cross-over regime (see
also~\cite{JF_Varenna}).
\subsection{Ground-state properties}
Previous numerical work mainly focused on the single-polaron problem. There
are two control parameters common in use, the first being the adiabaticity
ratio $\omega_0/t$.  In the adiabatic limit $\omega_0/t\to 0$, the motion of
the particle is affected by quasi-static lattice deformations, whereas in the
opposite anti-adiabatic limit $\omega_0/t\to\infty$, the lattice deformation
adjusts instantaneously to the carrier position.  The second parameter is the EP
coupling strength $\lambda=\ep/2\mathrm{D}t$, the ratio of the polaron
binding energy of an electron confined to a single site and the free
electron half-bandwidth in D dimensions.
\begin{figure}[t]
\includegraphics[width=.48\linewidth,clip]
{fehske_ahw_f2a.eps}\hspace*{0.5cm}
\includegraphics[width=.48\linewidth,clip]
{fehske_ahw_f2b.eps}\\[0.2cm]
\includegraphics[width=.48\linewidth,clip]
{fehske_ahw_f2c.eps}\hspace*{0.8cm}
\includegraphics[width=.45\linewidth,clip]
{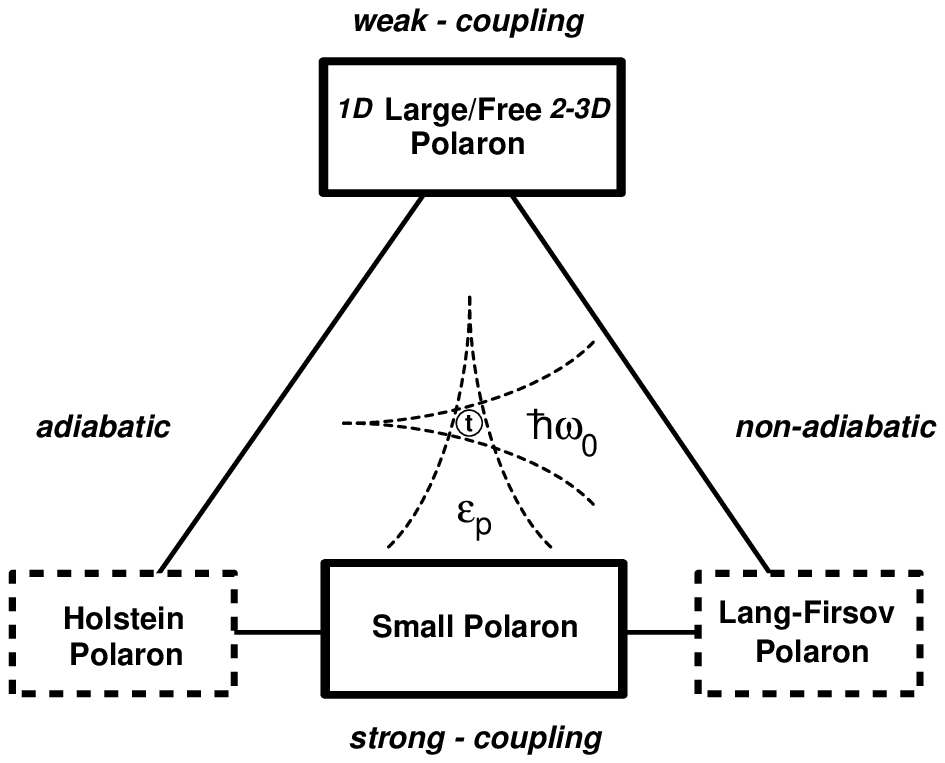}
\caption{Effective mass $m^*/m_0$ (data taken from~\cite{KTB02}),
  (normalised) electron-phonon correlation function $\chi_{0,j}= \langle
  n_0(b^\dagger_{0+j} + b^{}_{0+j})\rangle/2g\langle n_0\rangle $, and
  schematic ``phase diagram'' of the Holstein model.  Depending on the
  adiabaticity of the system, the cross-over regime is determined by the more
  stringent of the two conditions $\lambda \simeq 1$ and $g\simeq 1$. Thus
  starting from ``light'' ($\omega_0/t<1$) or ``heavy'' ($\omega_0/t>1$)
  electrons it is possible to understand the formation of small adiabatic
  ``Holstein'' or anti-adiabatic ``Lang-Firsov'' polarons as two limiting
  cases of a general picture.  }
\label{f:hpbr}
\end{figure}

Figure~\ref{f:hpbr} summarises selected ground-state characteristics of
small-polaron formation, showing results obtained by (variational)
ED~\cite{KTB02,WF98a} and DMRG~\cite{JW98a}. The transition to a polaron with large effective mass
takes place at $\lambda\simeq 1$, and is much sharper in higher
dimensions~\cite{KTB02}.  Adiabatic theory predicts an energy barrier
separating quasi-free (infinite radius) polarons from small-sized lattice
polarons in $\mathrm{D}>1$, but not in $\mathrm{D}=1$. However, the
(variational) ED data show that for $\omega_0>0$, the cross-over is
continuous in any dimension, i.e., the Holstein model does not exhibit a true
phase transition~\cite{KTB02,WF98a}, in accordance with the theorem of
Gerlach and L\"owen~\cite{GL91}.  In the anti-adiabatic regime, a further
parameter ratio, $g^2=\ep/\omega_0$, turns out to be crucial. Small-polaron
formation takes place if both $\lambda>1$ and $g^2>1$.  Note that the mass
enhancement notably deviates from the strong-coupling (perturbation theory)
result $\exp(g^2)$ at intermediate phonon frequencies~\cite{KTB02}.  The
cross-over from an only weakly phonon-dressed electron to a small polaron
also shows up very clearly in the correlation between electron- and 
phonon-displacements (not be equated with the polaron radius): At the critical
coupling, there is a strong enhancement of the on-site
correlations~\cite{WF98a,JW98a}.  Furthermore, $\chi_{0,j}/ \chi_{0,0}$
decays more rapidly as D increases, i.e., the surrounding phonons are
localised closer to the electron in higher dimensions~\cite{KTB02}.

The ground-state properties discussed so far constitute only one aspect of
polaron physics.  Of equal importance is the dynamical response of a
polaronic system to external perturbations.  For a moving (small) polaron
there is a perpetual exchange of momentum between the electron and the
deformation field~\cite{AR92a}.  Photoemission spectroscopy and inelastic
neutron scattering should therefore be able to detect the strong
interrelation of electron and phonon degrees of freedom. The strong mixing
of purely electronic states with lattice vibrational excitations should
be visible in the optical response as well. In the following sections, we
therefore present exact numerical results for electron/phonon spectral
functions and the optical conductivity in the framework of the Holstein
model.  A very efficient Chebyshev-expansion-based algorithm for the
calculation of such dynamical correlation functions is outlined
in~\cite{JF_Varenna} (for more details see \cite{WWAF05}).
\section{Photoemission spectra}\label{sec:photo}
Examining the dynamical properties of polarons, it is of particular interest
whether a QP-like excitation exists in the spectrum. This is probed by direct
(inverse) photoemission, where a bare electron is removed (added) from (to)
the many-particle system. The intensities (transition amplitudes) of these
processes are determined by the imaginary part of the retarded one-particle
Green's functions,
\begin{equation}\label{Gplusminus}
G^\pm(k,\omega) = 
\langle\langle c_k^\mp; c_k^\pm\rangle\rangle_\omega
=\lim_{\eta\to 0^+}\ \langle \psi_0 | c_k^\mp\,[\omega +\mathrm{i}\eta
\mp H]^{-1}\, c_k^\pm |\psi_0\rangle\,,
\end{equation}
i.e., by the wave-vector resolved spectral functions
\begin{equation}
\label{aspekt}
A^{\pm}(k,\omega) =-\frac{1}{\pi}\mathrm{Im}\,G^\pm(k,\omega)
=  \sum_{m} 
|\langle \psi_m^{\pm}|c_{k}^{\pm} 
|\psi_0^{}\rangle|^2 \,\delta [\,\omega\mp(E_m^{\pm}-E_0^{})]  
\end{equation}
with
$c_{k}^{+}=c_{k}^{\dagger}$ and $c_{k}^{-}=c_{k}^{}$.  These functions test
both the excitation energies $E_m^{\pm}-E_0^{}$ and the overlap of the ground
state $| \psi_0\rangle$ with the exact eigenstates $|\psi_m^{\pm}\rangle$ of
a $(N_{\rm e} \pm 1)$-particle system.  Hence, $G^+(k,\omega)$
[$G^-(k,\omega)$] describes the propagation of an additional electron [a
hole] with momentum $k$ [$-k$] and energy $\omega$. 
The electron spectral function of the single-particle Holstein model
corresponds to $N_{\rm e} = 0$, i.e., $A^-(k,\omega)\equiv 0$. 
$A(k,\omega)= A^+(k,\omega)+A^-(k,\omega)$  
can be determined, e.g., by cluster perturbation theory
(CPT)~\cite{SPP00,JF_Varenna}: We first calculate the Green's function
$G^c_{ij}(\omega)$ of a $N_c$-site cluster with open boundary conditions for
$i,j=1,\dots,N_c$, and then recover the infinite lattice by pasting 
identical copies of this cluster along the edges, treating the
inter-cluster hopping in first-order perturbation theory.

\begin{figure}[t]
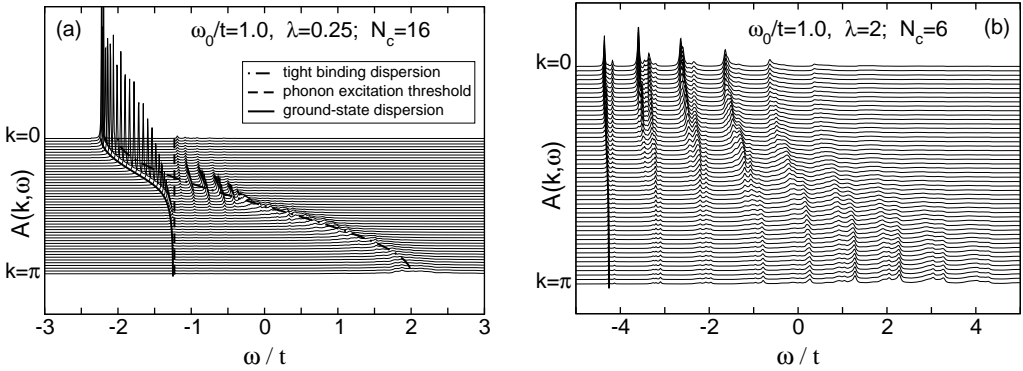

\includegraphics[width=.48\linewidth,clip]
{fehske_ahw_f3a.eps}\hspace*{0.5cm}
\includegraphics[width=.48\linewidth,clip]
{fehske_ahw_f3b.eps}
\caption{Spectral function of the 1D Holstein polaron calculated within CPT
  in the weak (a) and strong (b) non-adiabatic EP coupling regime. CPT
  is based on ED of a finite cluster with $N_c$ sites,  $M=7$
  ($\lambda=0.25$) and $M=25$ ($\lambda=2$) phonon quanta.}
\label{f:cpt_elekspe}
\end{figure}

Figure~\ref{f:cpt_elekspe} shows that at {\it weak coupling} (left panel),
the electronic spectrum is nearly unaffected for energies below the phonon
emission threshold. Hence, for the case considered here with $\omega_0$ lying
inside the bare electron bandwidth $4Dt$, the renormalised dispersion $E(k)$
follows the tight-binding cosine dispersion (lowered $\propto \ep$) up to
some $k_X$, where the dispersionless phonon intersects the bare electron
band. For $k>k_X$, electron and phonon states  ``hybridise'', and repel each
other, leading to the well-known band-flattening phenomenon~\cite{WF97}.  The
high-energy incoherent part of the spectrum is broadened $\propto \ep$, with
the $k$-dependent maximum again following the bare cosine dispersion.
\begin{figure}[b]
\begin{minipage}{0.5\linewidth}
\includegraphics[width=\linewidth,clip]
      {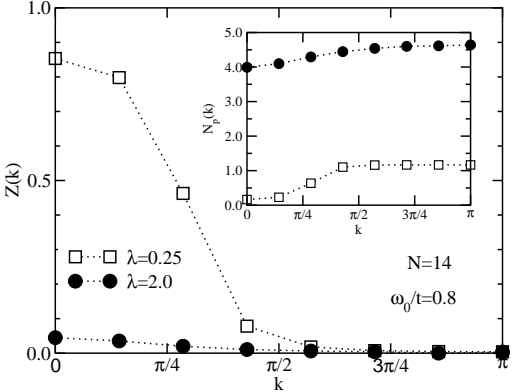}
\end{minipage}\hspace*{.5cm}
\begin{minipage}{0.47\linewidth}\vspace*{2.5cm}
  \caption{QP residue $Z(k)$ and mean phonon number $N_{\rm p}(k)$ (inset)
    obtained by ED at the allowed $k$ vectors of a 1D 14-site lattice with
    periodic boundary conditions. Note that $g^2 $ roughly gives the mean
    phonon number in the strong-coupling case.}
\label{f:zk_fk}
\end{minipage}
\end{figure}

To substantiate this interpretation, we compute the QP weight and mean
phonon number, 
\begin{equation}
\label{z_np}
Z(k)=|\langle \psi^+_k|c^\dagger_k|\psi_0^{}\rangle|^2\qquad\mbox{and}\qquad
N_{\rm p}(k)=\sum_i\langle \psi^+_k|(b_i^\dagger+b_i^{})|\psi_0^{}\rangle\,,
\end{equation}
for each $k$-sector (see Fig.~\ref{f:zk_fk}).  The $k$-dependent $Z$-factor
can be taken as a measure of the ``electronic contribution'' to the QP.  For
weak EP coupling, we have $Z(k\ll k_X)\lesssim 1$ [$Z(k>k_X)\ll 1$],
reflecting the electronic [phononic] character of the states at the band
centre [band edge], which are basically zero-phonon [one-phonon] states (see
Fig.~\ref{f:zk_fk}, inset).  Increasing the EP coupling, a strong mixing of
electron and phonon degrees of freedom takes place, whereby -- forming a
small polaron -- both quantum objects cease to exist independently.  As
expected, this leads to a significant suppression of the (electronic) QP
residue $Z(k)$ for all~$k$.  Now the states $|\psi_0^{}\rangle$,
$|\psi_k^\pm\rangle$ are multi-phonon states, and the polaronic QP is heavy
because it has to drag with it a large number of phonons in
its phonon cloud (cf. Fig.~\ref{f:zk_fk}, inset).

The inverse photoemission spectrum in the {\it strong-coupling case} is shown
in the right panel of Fig.~\ref{f:cpt_elekspe}. First, we observe all
signatures of the famous polaronic band-collapse, where a well-separated,
narrow (i.e., strongly renormalised), coherent QP band is formed at
$\omega\simeq -\ep$.  If we had calculated the polaronic instead of the
electronic spectral function~\eqref{aspekt}, nearly all spectral weight would
reside in the coherent part, i.e., in the small-polaron band~\cite{FLW97}.
In contrast, $Z(k)$, defined by~\eqref{z_np}, is extremely small and
approaches the strong-coupling result $Z=\exp(-g^2)$ for $\lambda\,,g^2 \gg
1$.  Note that the inverse effective mass $m^*/m_0$ and $Z(k)$ differ if the
self-energy is strongly $k$-dependent. This discrepancy has its
maximum in the intermediate-coupling regime for 1D systems, but vanishes in
the limit $\lambda\to \infty$ and, in any case, for $D=\infty$~\cite{KTB02}.
The incoherent part of the spectrum is split into several sub-bands separated
in energy by $\omega_0$, corresponding to excitations of an electron and
  one or more phonons (Fig.~\ref{f:cpt_elekspe}).

Finally, let us emphasise that for all couplings, the lowest-lying band in
$A(k,\omega)$ almost perfectly coincides with the coherent polaron
band-structure (solid line in Fig.~\ref{f:cpt_elekspe}) obtained by
variational ED~\cite{KTB02}. This method has been shown to give very accurate
results for the infinite system, also underlining the high precision of our
CPT approach.
\section{Phonon spectra}
\begin{figure}[t]
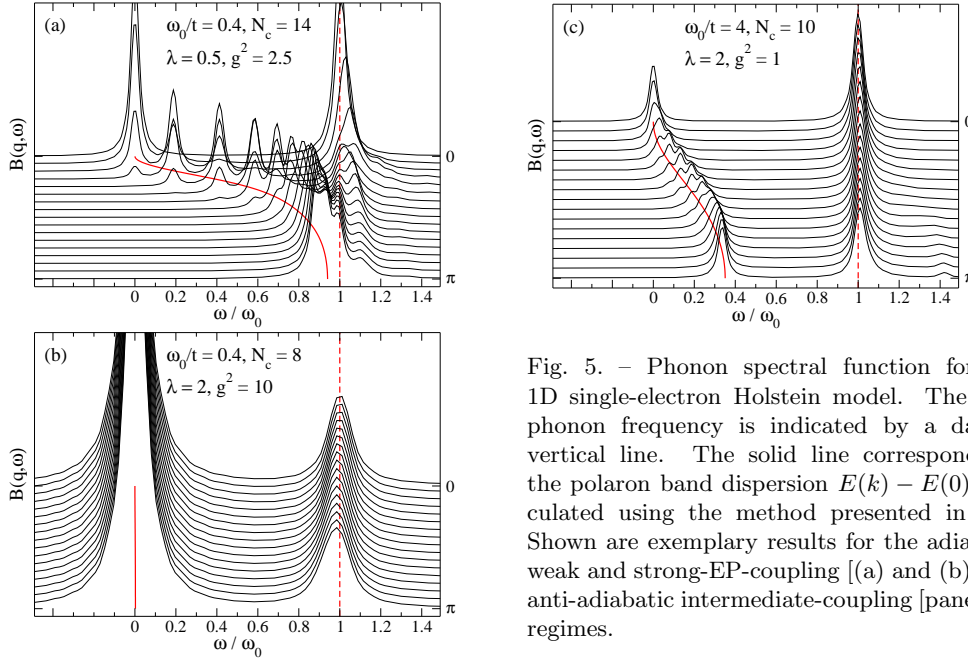

\begin{minipage}{0.49\linewidth}
\includegraphics[width=.9\linewidth,clip]
{fehske_ahw_f5a.eps}\\
\includegraphics[width=.9\linewidth,clip]
{fehske_ahw_f5b.eps}
\end{minipage}\hspace*{0.2cm}
\begin{minipage}{0.49\linewidth}
\includegraphics[width=.9\linewidth,clip]
{fehske_ahw_f5c.eps}
\caption{Phonon spectral function for the 1D single-electron Holstein model.
  The bare phonon frequency is indicated by a dashed vertical line.  The
  solid line corresponds to the polaron band dispersion $E(k)-E(0)$,
  calculated using the method presented in~\cite{KTB02}.  Shown are
  exemplary results for the adiabatic weak and strong-EP-coupling [(a) and
  (b)] and anti-adiabatic intermediate-coupling [panel (c)] regimes.}
\label{f:phonon_spect}
\end{minipage}
\end{figure}
Next we show that the phonon spectra provide additional useful information
concerning the polaron dynamics. For this purpose, we calculate the $T=0$
spectral function $B(q,\omega)$, which is related to the phonon Green's function $D(q,\omega)$ for $\omega>0$ by
\begin{equation}
\label{bspekt}
B(q,\omega)=-\frac{1}{\pi}\mathrm{Im}\,D(q,\omega)\,,\quad
D(q,\omega)=2 \omega_0\langle\langle x_q;x_{-q}\rangle\rangle_\omega\,, 
\end{equation}
where $x_q=N^{-\mbox{\small $\frac{1}{2}$}}\sum_j x_j \mbox{e}^{-{\rm i}j
  q}$ and $x_j=(b^\dagger_j+b^{}_j)/\sqrt{2\omega_0}$.  Again we employ a
cluster approach, determining first the cluster phonon Green's function
$D_{ij}^{c}(\omega)$, and afterwards, as in CPT, constructing the energy
and momentum dependent Green's function of the infinite lattice from
$D(q,\omega) = \frac{1}{N_c} \sum_{i,j=1}^{N_c} D^{c}_{ij}(\omega) \text{
  e}^{-{\rm i}q(i-j)}$.  Since for the Holstein model the
bare phonon Green's function 
$D^0(\omega) = 2\omega_0/(\omega^2-\omega_0^2)$ is $k$-independent, 
this cluster expansion is identical to replacing the full real-space Green's function $D_{ij}$ by $D^c_{ij}$. 

Figures~\ref{f:phonon_spect}(a) and~(b) show the evolution of the phonon
spectrum with increasing EP interaction in the adiabatic case
($\omega_0/t=0.4$). For $\lambda=0.5$, there is a dispersive low-energy
absorption reflecting the polaron band dispersion, as demonstrated by the
comparison with variational-ED data for $E(k)$. If the weakly renormalised
electron band intersects the (dispersionless) bare phonon excitation at some
$q_Y$, level repulsion occurs and we observe two adjacent
absorption features.  At larger EP couplings, the polaron band separates
completely from the bare phonon signature until, in the extreme
strong-coupling limit, two almost flat absorption bands emerge, corresponding
to the lowest (small) polaron band and the first excited band, separated
by a one-phonon excitation. Both signals are strong because of the large
``phonon content'' in these states.

As stated above, in the anti-adiabatic regime, the small-polaron cross-over
is determined by the ratio $g^2$, and occurs at about $g^2=1$.  For
$g^2\ll 1$, nearly the whole spectral weight resides in the bare phonon peak
(if $\omega_0>4Dt$).  The phonon spectrum near the transition point is shown
in Fig.~\ref{f:phonon_spect}(c) for $\omega_0/t=4$.  We detect a clear
signature of the small-polaron band with renormalised width $W\simeq
1.5t$ -- a factor of about ten larger than in the adiabatic case ($\lambda=1$,
$\omega_0/t=0.4$; note that the excitation energy in
Fig.~\ref{f:phonon_spect} is plotted in units of $\omega_0$). The
dispersionless excitation at $\omega_0$ is obtained by adding one phonon with
momentum $q$ to the $k=0$ ground state.  Above this pronounced peak, we find
an ``image'' of the lowest polaron band -- shifted by $\omega_0$ -- with
extremely small spectral weight, hardly resolved in
  Fig.~\ref{f:phonon_spect}.
\section{Optical response}
\begin{figure}[b]
 \includegraphics[width=0.98\linewidth,clip]
      {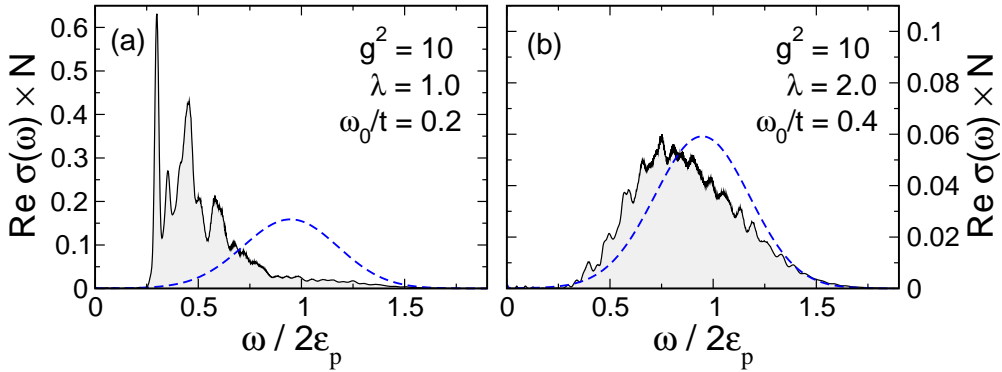}
      \caption{Optical conductivity (in units of $\pi e^2 t^2$) of the 1D
        Holstein polaron at $T=0$ compared to the analytical small-polaron
        result (dashed line). ED data are for a system with $N=6$ 
        sites and 45 phonons; $\sigma_0$ is determined to give the same
        integrated spectral weight as $\mbox{Re}\, \sigma(\omega>0)$.}
   \label{acsigma_kpm_reik}
  \end{figure}
We apply the ED-KPM scheme outlined in~\cite{JF_Varenna,WWAF05,SWWAF05} to
calculate the optical absorption of the 1D Holstein polaron.  The results for
the (regular) real part of the conductivity,
\begin{equation}\label{sigma}
  \mbox{Re}\,\sigma(\omega) = \frac{\pi}{\omega N} \sum_{E_m>E_0} 
  |\langle \psi_m|\hat{\jmath}|\psi_0\rangle |^2\ 
\delta[\omega - (E_m - E_0)]
\end{equation}
(here $\hat{\jmath} = - \mbox{i} e t\sum_{i}(c_{i}^{\dagger} c_{i+1}^{} -
c_{i+1}^{\dagger} c_{i}^{})$ is the current operator), and possible
deviations from established polaron theory are important for relating theory
to experiment.  The standard description of small polaron
transport~\cite{Emi93} yields for the $T=0$ ac conductivity
\begin{equation}\label{sigma_sc}
\mbox{Re}\, \sigma (\omega) = 
  \frac{\sigma_0}{\sqrt{\ep\omega_0}} \frac{1}{\omega} 
  \exp \left[- \frac{(\omega-2 \ep)^2}{4 \ep \omega_0}\right]\,.
\end{equation} 
For sufficiently strong coupling this formula predicts a weakly asymmetric
Gaussian absorption peak centred at twice the polaron binding energy.

Figure~\ref{acsigma_kpm_reik} shows $\mbox{Re}\, \sigma(\omega)$ when polaron
formation sets in (a), and above the transition point (b).
For $\lambda=2$ and $\omega_0/t=0.4$, i.e., at rather large EP coupling but
not in the extreme small-polaron limit, we find a pronounced maximum in
the low-temperature optical response, which, however, is located somewhat
below $2\ep=2g^2\omega_0$, the value for small polarons at $T=0$. At
the same time, the line-shape is more asymmetric than in small-polaron
theory, with a weaker decay at the high-energy side, fitting even better to
experiments on standard polaronic materials such as TiO$_2$~\cite{KMF69}.  At
smaller couplings, significant deviations from a
Gaussian-like absorption are found, i.e., polaron motion is not adequately
described as hopping of a self-trapped carrier almost localised on a single
lattice site.
\section{Many-polaron problem}
Finally, we address the important issue how the character of the QPs
of the system changes with carrier density. 
While for very strong EP coupling no
significant changes are expected due to the existence of rather independent
small (self-trapped) polarons with negligible residual interaction, a
density-driven cross-over from a state with large polarons to a metal with
weakly dressed electrons may occur in the intermediate-coupling 
regime~\cite{HNLWLF05}. This
problem has recently been investigated experimentally by optical
measurements on \chem{La_{2/3}(Sr/Ca)_{1/3}MnO_3} films~\cite{HMDLK04}.

In the Holstein model, the above-mentioned density-driven
transition from large polarons to weakly EP-dressed
electrons is expected to be possible only in 1D, where large polarons exist
at weak and intermediate coupling. The situation is different for
Fr\"{o}hlich-type models~\cite{Fr54,AK99,DT01} with long-range EP
interaction, in which large-polaron states exist even for strong coupling and
in $\mathrm{D}>1$.
\begin{figure}[t]
\includegraphics[width=.48\linewidth]
{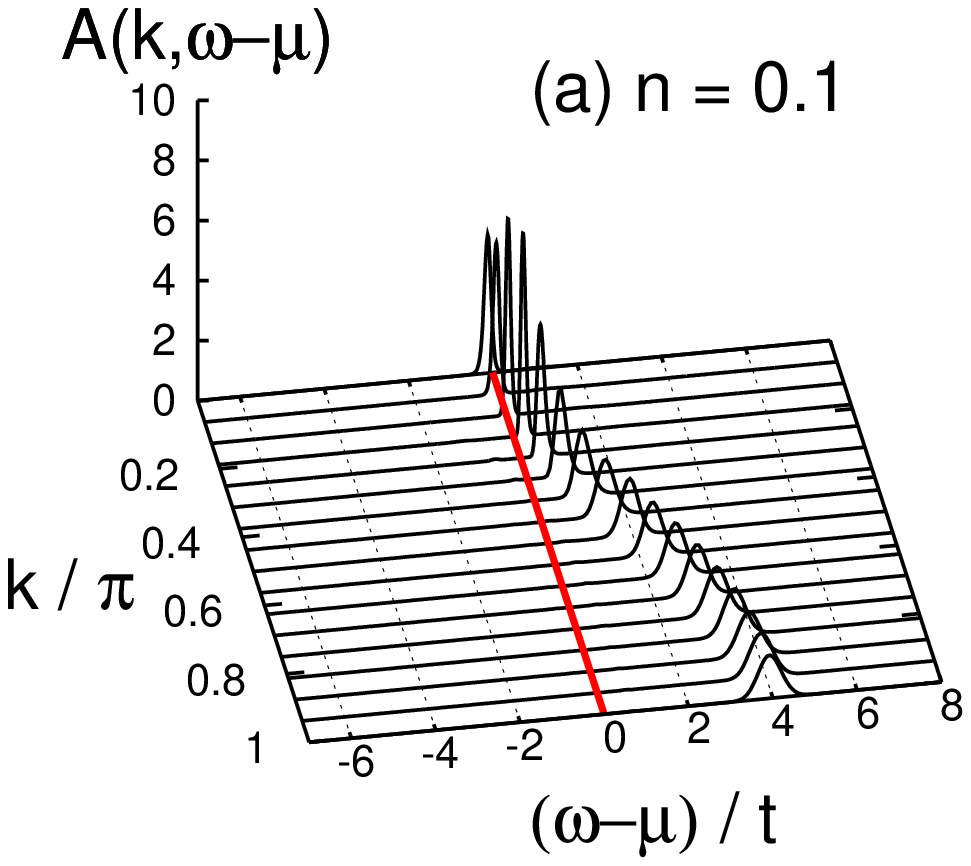}
\includegraphics[width=.48\linewidth]
{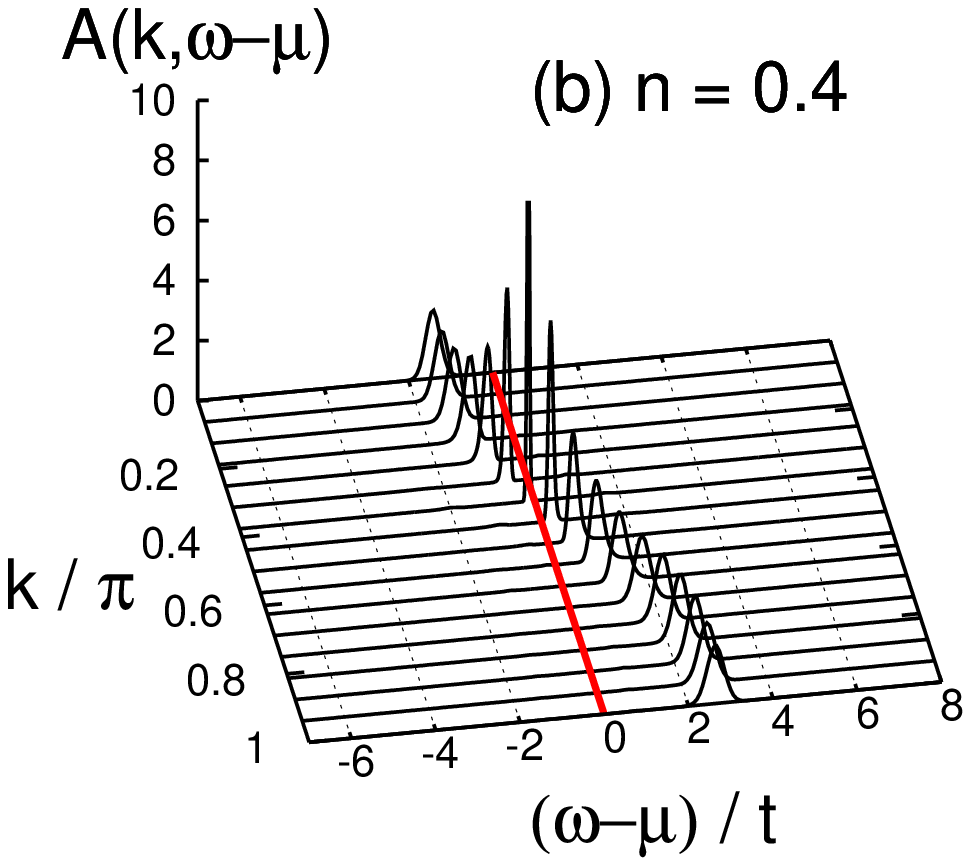}\\[0.2cm]
\includegraphics[width=.48\linewidth]
{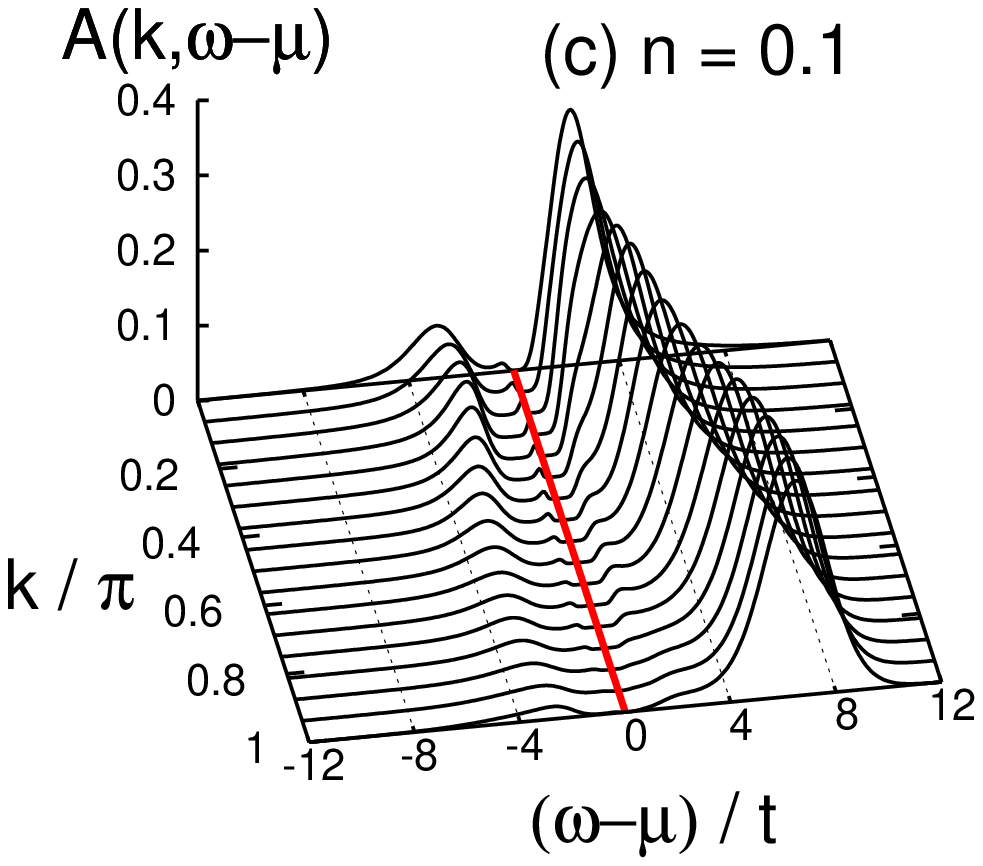}
\includegraphics[width=.48\linewidth]
{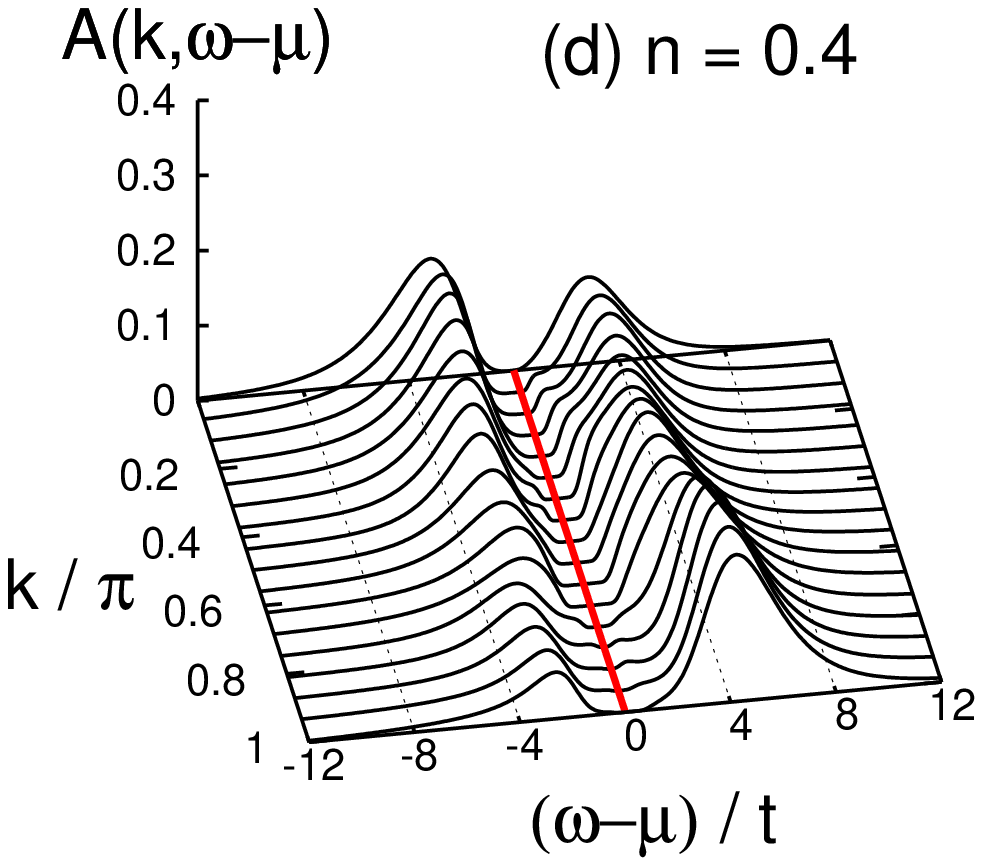}
\caption{One-electron spectral function $A(k,\omega-\mu)$ of the 1D spinless
  Holstein model from QMC~\cite{HNLWLF05} for low ($n=0.1$)
  and high ($n=0.4$) carrier densities at weak [(a),(b); $\lambda=0.1$] and
  strong [(c),(d); $\lambda=2$] EP coupling, $\omega_0/t=0.4$, $N=32$,
  and $t/k_{\rm B}T=8$ (inverse temperature). Note
  that QMC cannot resolve features with very small spectral weight.}
\label{f:results_ak_QMC}
\end{figure}

To set the stage, Fig.~\ref{f:results_ak_QMC} shows the evolution of the
one-electron spectral function $A(k,\omega-\mu)$ with increasing electron
density $n$ in the {\it weak- and strong-coupling limiting cases}.  In the
former [Fig.~\ref{f:results_ak_QMC}(a),(b)], the spectra bear a close
resemblance to the free-electron case for all $n$, i.e., there is a strongly
dispersive band running from $-2t$ to $2t$, which can be attributed to weakly
dressed electrons with an effective mass close to the non-interacting value.
As expected, the height (width) of the peaks increases (decreases)
significantly in the vicinity of the Fermi momentum, which is determined by
the crossing of the band with the chemical potential $\mu$.  In the opposite
strong-coupling limit [Fig.~\ref{f:results_ak_QMC}(c),(d)], the spectrum
exhibits an almost dispersionless coherent polaron band at $\mu$ (QMC, due to
the use of maximum entropy, has problems resolving such extremely weak
signatures).  Besides, there are two incoherent features located above and
below the chemical potential, broadened $\propto \ep$, reflecting
phonon-mediated transitions to high-energy electron states.  At $n=0.4$, the
photoemission spectrum for $k<\pi/2$ becomes almost symmetric to the inverse
photoemission spectrum for $k>\pi/2$ and already reveals the gapped structure
expected at $n=0.5$ due to the Peierls transition (charge-density-wave
formation).  The
most important point, however, is the {\it clear separation} 
of the coherent band 
from the incoherent parts even at large $n$.  This indicates that small
polarons are well-defined QPs in the strong-coupling regime, even at high
carrier density.

\begin{figure}[t]\hspace*{-.2cm}
\includegraphics[width=.6\linewidth]
{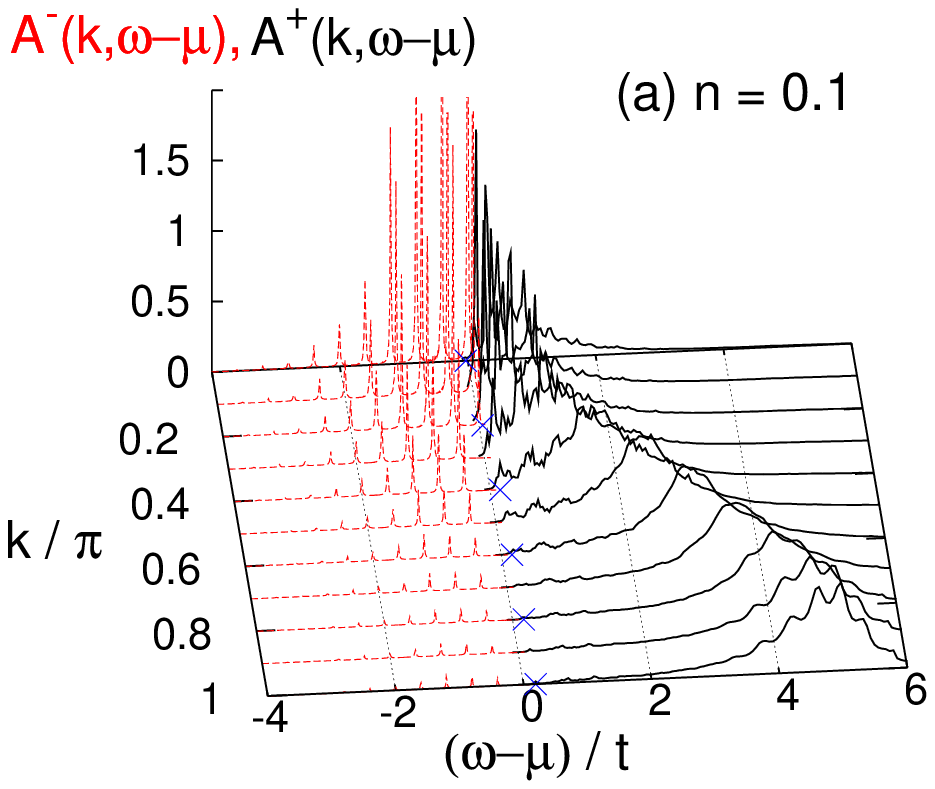}\hspace*{-1.1cm}
\includegraphics[width=.6\linewidth]
{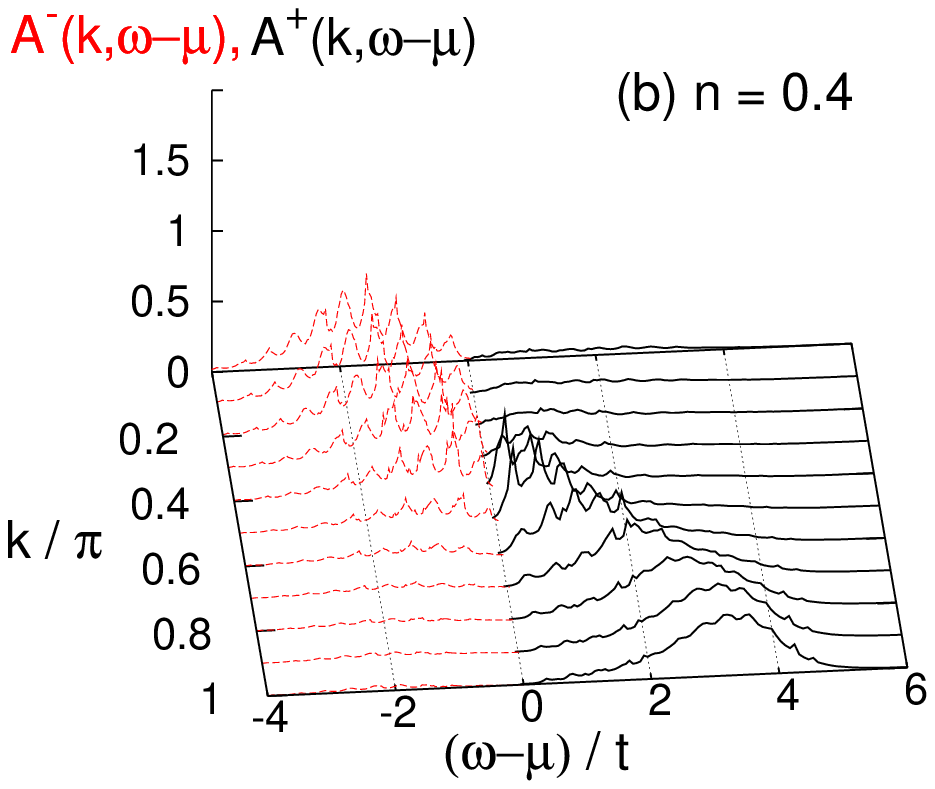}
\caption{Single particle spectral functions $A^-(k,\omega)$ (red dashed
  lines) and $A^+(k,\omega)$ (black solid lines) for different band fillings
  $n$ at $\omega_0/t=0.4$ and $\lambda=1$ (CPT results with $N_c=10$).
Blue crosses track the small-polaron band obtained by ED.}
\label{f:manypol_akw_density}
\end{figure}

Figure~\ref{f:manypol_akw_density} displays the inverse photoemission
[$A^+(k,\omega)$] and photoemission spectra [$A^-(k,\omega)$] at
{\it intermediate EP coupling strength}, determined by CPT at $T=0$. 

At $n=0.1$ we can also identify a (coherent) 
polaron band (blue crosses; cf. also the ED data in~\cite{HNLWLF05}).
The Fermi energy $E_\mathrm{F}=\mu(T\to 0)$ is located within this signature.
The large-polaron band has small electronic spectral weight especially
away from  $E_\mathrm{F}$ and flattens as an effect of the EP coupling 
at large $k$ (see Sec.~\ref{sec:photo}). 
Below this band, there exist equally spaced phonon satellites, 
reflecting the Poisson distribution
of phonons in the ground state.  Above $E_\mathrm{F}$ there is a broad
dispersive incoherent feature whose maximum follows 
closely the dispersion relation of free particles.

As $n$ increases, a well-separated coherent polaron band 
can no longer be identified.
At about  $n\simeq 0.3$ the deformation clouds of the 
(large) polarons start to overlap leading to a mutual
(dynamical) interaction between the particles. Increasing the carrier
density further,
the polaronic QPs dissociate, stripping their phonon cloud. 
Now diffusive scattering of electrons and phonons 
seems to be the dominant interaction mechanism. 
As a result both, the phonon peaks in $A^-(k,\omega)$ and 
the incoherent part of $A^+(k,\omega)$ are washed out, the spectra
broaden and ultimately merge into a single wide band~\cite{HNLWLF05,HWAF05}.
As can be seen from Fig.~\ref{f:manypol_akw_density} (b), the 
incoherent excitations lie now arbitrarily close to the Fermi level. 
These are significant differences to the nearly free electron and
small polaron spectra shown for the same carrier density in 
Figs.~\ref{f:results_ak_QMC}~(b) and (d), respectively.  
Obviously the low-energy physics of the 
system can no longer be described by small-polaron theory.
\section{Summary and open problems}
In this contribution, we have reviewed ground-state- and, most notably,
spectral properties of Holstein polarons by means of quasi-exact numerical
methods such as (variational) Lanczos diagonalisation, a kernel polynomial
expansion technique, cluster perturbation theory, and quantum Monte Carlo.
Our numerical approaches yield unbiased results in all
parameter regimes, but are of particular value in the non-adiabatic
intermediate-coupling regime, where perturbation theories and other
  analytical techniques fail. The data presented for the (inverse)
photoemission, phonon- and optical spectra show that electron and phonon
excitations become intimately, dynamically connected in the process of
polaron formation.

Although we have now achieved a rather complete picture of the single
(Holstein) polaron problem (perhaps dispersive phonons, longer-ranged EP
interaction, finite temperature and disorder effects deserve closer
attention), the situation is discontenting in the case of a finite carrier
density. Here electron-phonon coupling competes with sometimes strong
electronic correlations as in, e.g., unconventional 1D metals, quasi-1D MX
chains, quasi-2D high-$T_c$ superconductors, 3D charge-ordered nickelates, or
bulk colossal magneto-resistance manganites. The corresponding microscopic
models contain (extended) Hubbard, Heisenberg or double-exchange terms,
and maybe also a coupling to orbital degrees of freedom, so that they can
hardly be solved even numerically with the same precision as the Holstein
model. Consequently, the investigation of these materials and models will
definitely be a great challenge for solid-state theory in the near future.

\acknowledgments
We would like to thank J.~Bon\v{c}a, D.~Ihle, J.~Loos, S.~A.~Trugman,
G.~Schubert and A.~Wei{\ss}e for valuable discussions.

\end{document}